\documentstyle{article}

\newcommand{\bea}{\begin{eqnarray}}
\newcommand{\eea}{\end{eqnarray}}

\begin{document}
\baselineskip = 12pt

\title{The Origin of Structures in Generalized Gravity}

\author{Jai-chan Hwang \\
        {\sl Department of Astronomy and Atmospheric Sciences} \\
        {\sl Kyungpook National University, Taegu, Korea}
        }

\date{}
\maketitle
\parindent = 2em

\begin{abstract}

In a class of {\it generalized gravity theories} with general couplings 
between the scalar field and the scalar curvature in the Lagrangian, 
we can describe the {\it quantum generation} and the 
{\it classical evolution} of both the scalar and tensor structures 
in a simple and unified manner.
An accelerated expansion phase based on the generalized gravity
in the early universe drives microscopic quantum fluctuations inside 
a causal domain to expand into macroscopic ripples in the spacetime metric 
on scales larger than the local horizon.  
Following their generation from quantum fluctuations, the ripples in 
the metric spend a long period outside the causal domain.  
During this phase their evolution is characterized by their 
{\it conserved} amplitudes. 
The evolution of these fluctuations may lead to the observed large scale 
structures of the universe and anisotropies in the cosmic microwave
background radiation.

\end{abstract}

\subsection*{1. Historical perspective}

The classical evolution of structures in an expanding universe was first 
analyzed in the context of General Relativity in a classic study by 
E. M. Lifshitz in 1946 \cite{Lifshitz}.
The theoretical introduction of an accelerated expansion (inflation) 
phase in the early universe \cite{Inflation} enables us to draw a coherent 
picture of the origin of the large scale structures in the universe:
The ever-present microscopic vacuum quantum fluctuations become macroscopic 
during this acceleration phase and can subsequently develop into the observed 
large-scale structures.

Soon after the introduction of the field equation of General Relativity 
by A. Einstein in 1915 \cite{Einstein} and its immediate action formulation 
by D. Hilbert \cite{Hilbert}, H. Weyl, W. Pauli, and A. S. Eddington \cite{Weyl}
considered modifications of the theory involving the addition of 
general curvature combinations to the action.
Mach's principle motivated the Brans-Dicke gravity \cite{Brans-Dicke}, 
and the notion of spontaneous symmetry breaking led to the idea of 
induced gravity \cite{induced-gravity}.
Generalized forms of Einstein gravity almost always appear in any reasonable
attempt to understand the quantum aspects of the gravity theory, and
also naturally appear in the low energy limits of diverse attempts to 
unify gravity with other fundamental forces \cite{QFCS}.
Consequently, it seems increasingly likely that the early stages of the 
evolution of the universe were governed by a gravity law more general than 
Einstein's theory.

\subsection*{2. Three stages}

Current theoretical attempts to explain the origin of large scale structure
in the universe can be described in three stages.
We consider an expanding universe model modified by an accelerated expansion 
era in the early stage.
The first stage is the {\it quantum generation stage}: structural seeds 
are generated from the ever present microscopic quantum fluctuations 
of the fields and the metric during the acceleration era.
Due to the acceleration of cosmological expansion, quantum fluctuations 
residing in a causal domain are pushed outside the local horizon defined by a 
light travel distance.
The second is the {\it classical evolution stage}: the magnified 
structures of quantum-mechanical origin become classical as their sizes become 
bigger than the local horizon.
The evolution of fluctuations during this superhorizon stage is kinematic and 
is described by linear fluctuation theory based on classical relativistic 
gravity.
In the later parts of this stage, as the expansion decelerates, 
the horizon scale overtakes the scales of the observationally relevant 
fluctuations, and makes these structures visible to the observer.
The third stage is the {\it nonlinear evolution stage}: the smaller
scale scalar type structures become nonlinear.
Due to the extremely low level anisotropy of the cosmic microwave
background radiation \cite{COBE}, nonlinear evolution is expected to 
start only on scales well within the horizon scale of the matter 
dominated era governed by Einstein gravity.
Thus, the nonlinear evolution stage is usually handled in the Newtonian context.
In this paper, we will address the role played by  a class of 
generalized gravity theories in the first two stages and the unified 
perspective we can derive.

\subsection*{3. Generalized gravity and the perturbed universe}

We consider a class of generalized gravity theories with an action
\bea
   S = \int d^4 x \sqrt{-g} \left[ {1 \over 2} f (\phi, R)
       - {1\over 2} \omega (\phi) \phi^{;a} \phi_{,a} - V(\phi)
       + \; L_m \right],
   \label{Action-GGT}
\eea
where $f$ is a general algebraic function of the scalar (dilaton) field $\phi$ 
and the scalar curvature $R$, and $\omega$ and $V$ are general functions of 
$\phi$; $L_m$ is an additional matter part of the Lagrangian.  Equation
(\ref{Action-GGT}) includes the following generalized gravity theories as 
subsets:
(a) generally coupled scalar fields,
(b) generalized scalar tensor theories which include the Brans-Dicke theory,
(c) induced gravity,
(d) the low energy effective action of string theory,
(e) $f(R)$ gravity, etc.
Einstein gravity is a case with $f = R$ and $\omega = 1$.
Each gravity theory in (a-e) without $L_m$ corresponds to a single 
component system; the gravity theory in (\ref{Action-GGT}) without $L_m$
in general corresponds to a two field system.

As the background universe, we consider a spatially homogeneous and isotropic
metric with a vanishing spatial curvature and cosmological constant.
Since the structures in the first two stages are assumed to be linear,
they can be handled by perturbations of the model universe.
The metric of the perturbed universe, including {\it the most general}
scalar, vector and tensor perturbations, can be written as
\bea
   d s^2 
   &=& - \left( 1 + 2 \alpha \right) d t^2
       - a^2 \left( \beta_{,\alpha} + B_\alpha \right) d t d x^\alpha 
   \nonumber \\
   & & + \; a^2 \Big[ \delta_{\alpha\beta} \left( 1 + 2 \varphi \right) 
       + 2 \gamma_{,\alpha|\beta} + 2 C_{(\alpha|\beta)}
       + 2 C_{\alpha\beta} \Big] d x^\alpha d x^\beta.
   \label{metric}
\eea
A scalar structure is characterized by $\alpha ({\bf x}, t)$,
$\beta ({\bf x}, t)$, $\varphi ({\bf x}, t)$, and $\gamma ({\bf x}, t)$; 
the transverse $B_\alpha ({\bf x}, t)$ and $C_\alpha ({\bf x}, t)$, 
and the transverse-trace-free $C_{\alpha\beta} ({\bf x}, t)$ describe 
vector and tensor structures, respectively (the perturbed order metric
variables have 10 degrees of freedom).
When we consider a perturbed spacetime since we are dealing with 
two metric systems (one is a perturbed metric and the other is a fictitious 
unperturbed metric) we have the freedom of gauge choices.
Due to the homogeneity of the background model, without losing generality,
we can choose the spatial gauge conditions $\gamma \equiv 0 \equiv C_\alpha$ 
(thus fixing three degrees of freedom) which completely fix the spatial
gauge degrees of freedom.
Under this spatial gauge condition all the remaining scalar perturbation 
variables are {\it spatially} gauge invariant, and the remaining vector 
perturbation variable is also gauge invariant;
tensor perturbation variables are naturally gauge invariant.
The remaining temporal gauge condition with one degree of freedom
only affects the scalar perturbation.
In the gauge ready method, this temporal gauge condition can be used as 
an advantage in handling the scalar type perturbation by choosing the
gauge condition according to the mathematical convenience of each
individual problem \cite{PRW}.
Except for the synchronous gauge condition which fixes $\alpha \equiv 0$
the rest of the fundamental temporal gauge condition completely fixes
the temporal gauge mode, thus each variable in such gauge condition
corresponds to a gauge invariant combination of the considered variable 
and the variable used in the gauge condition; see below (\ref{GI}).

After fixing the functional forms of $f (\phi, R)$, $\omega (\phi)$, 
$V (\phi)$, and the equation of state, the equations for the background 
will lead to a solution for the cosmic scale factor $a(t)$. 
Due to the high symmetry in the background, all three types of perturbations 
{\it evolve independently of each other}.
A {\it vector perturbation} is trivially described by a conservation 
of the angular momentum: for a vanishing anisotropic stress we have
$a^3 (\mu + p) \cdot a \cdot v_\omega \sim$ constant in time, where
$\mu (t)$, $p (t)$, and $v_\omega ({\bf x}, t)$ are the background energy 
density and pressure, and the vorticity part of the matter velocity in $L_m$. 
Remarkably, the generalized nature of the gravity does not affect this result
which is valid even considering the Ricci-curvature square term in the action,
see \cite{vorticity}.

\subsection*{4. The classical evolution of scalar and tensor structures}

{}For the scalar field we let
$\phi ({\bf x}, t) = \phi (t) + \delta \phi ({\bf x}, t)$.
When we consider a scalar perturbation the following 
gauge invariant 
combination plays an important role
\bea
   \delta \phi_\varphi \equiv \delta \phi - {\dot \phi \over H} \varphi
       \equiv - {\dot \phi \over H} \varphi_{\delta \phi}.
   \label{GI}
\eea
$\delta \phi_\varphi$ is the same as $\delta \phi$ in the uniform-curvature
gauge ($\varphi \equiv 0$), and $\varphi_{\delta \phi}$ is the same
as $\varphi$ in the uniform-field gauge ($\delta \phi \equiv 0$); for the
gauge transformation propery of each variable see \cite{PRW}.
The perturbed action to the second order in the perturbation variables
can be arranged in a remarkably simple and unified form 
(for derivation, see \cite{H-CT,H-GW,Mukhanov})
\bea
   \delta^2 S = {1 \over 2} \int a^3 Q \left( \dot \Phi^2
       - {1 \over a^2} \Phi^{|\gamma} \Phi_{,\gamma} \right) dt d^3 x, 
   \label{perturbed-action}
\eea
where for scalar and tensor perturbations, respectively, we have
($F \equiv \partial f/ \partial R$)
\bea
   \Phi = \varphi_{\delta \phi}, \quad
       Q = { \omega \dot \phi^2 + {3 \dot F^2 \over 2 F}
       \over \left( H + {\dot F \over 2 F} \right)^2 } \; ; \qquad 
       \Phi = C^\alpha_\beta, \quad Q = F.
   \label{Phi-Q}
\eea
The non-Einstein nature of the theory is present in the parameter $Q$.
The equation of motion becomes
\bea
   {1 \over a^3 Q} (a^3 Q \dot \Phi)^\cdot - {1 \over a^2} \nabla^{2} \Phi = 0.
\eea
This has a general large scale solution 
\bea
   \Phi = C - D \int_0^t {dt \over a^3 Q},
   \label{LS-sol}
\eea
where $C ({\bf x})$ and $D ({\bf x})$ are the integration constants for 
the growing and decaying modes, respectively.
This solution is valid for general $V(\phi)$, $\omega(\phi)$, and 
$f(\phi,R)$, and expresses perturbation evolution in 
a remarkably simple unified form; 
for the scalar type perturbation these results are valid for a single 
component subclass of (\ref{Action-GGT}) in (a-e) without $L_m$, 
whereas for the tensor type perturbation they are valid for 
the general action in (\ref{Action-GGT}).
It is noteworthy that the growing mode of $\Phi$ 
(thus, $\varphi_{\delta \phi}$ and $C_{\alpha\beta}$) 
is conserved in the large scale limit 
{\it independently} of the specific gravity theory under consideration.
It follows that the classical evolutions of very large scale perturbations
are characterized by conserved quantities.
These are {\it conserved even under the chaging gravity theories} as long as
the gravity theories belong to (a-e) for the scalar perturbation, and
for general theory in (\ref{Action-GGT}) for the gravitational wave.
[This conserved behavior also applies for sufficiently large scale
perturbations during the fluid eras in Einstein gravity models; 
in the fluid era the defining criteria for considering a
perturbation to be large scale are the Jeans scale 
(sound horizon) for a scalar structure and the visual horizon for 
a gravitational wave, \cite{PRW}.]  
The integration constant $C({\bf x})$ {\it encodes the information 
about the spatial structure} of the growing mode.
Thus, in order to have information about large scale structure,
we need information about $\Phi = C({\bf x})$
which must have been generated in some early evolutionary 
stage of the universe.

\subsection*{5. Quantum generations}

In order to handle the quantum mechanical generations of scalar 
structures and gravitational waves, we regard the perturbed parts
of the metric and matter variables as Hilbert space operators, $\hat \Phi$.
Having the perturbed action in (\ref{perturbed-action})
the process of quantization and the derivation of quantum fluctuations
are straightforward.
The correct normalization of the equal time commutation relation
follows from (\ref{perturbed-action}) as
\bea
   [ \hat \Phi ({\bf x},t), \dot {\hat \Phi} ({\bf x}^\prime, t) ]
       = {i \over a^3 Q} \delta^3 ({\bf x} - {\bf x}^\prime).
   \label{commutation}
\eea
[In the quantization of the gravitational wave we need to 
take into account of the two polarization states properly; 
we ignore this minor complication, see \cite{GW,H-GW}.]

{}{\it For} $a \sqrt{Q} \propto \eta^q$ ($d\eta \equiv dt/a$) we have an exact 
solution for the mode function
\bea
   \Phi_{\bf k} (\eta) = {\sqrt{ \pi |\eta|} \over 2 a \sqrt{Q}} 
       \Big[ c_1 ({\bf k}) H_\nu^{(1)} (k|\eta|)
       + c_2 ({\bf k}) H_\nu^{(2)} (k|\eta|) \Big], \quad
       \nu \equiv {1 \over 2} - q,
   \label{Phi-k-sol}
\eea
where according to (\ref{commutation}) we have
$|c_2 ({\bf k})|^2 - |c_1 ({\bf k})|^2 = 1$; the freedom in $c_1$ and $c_2$ 
indicates the dependence on the vacuum state.
The power spectrum based on the vacuum expectation value is
\bea
   & & {\cal P}^{1/2}_{\hat \Phi} ({\bf k}, \eta)
       = \sqrt{k^3 \over 2 \pi^2} | \Phi_{\bf k} |.
   \label{P-vac-def}
\eea
In the large scale limit we have, for $\nu \neq 0$ and $\nu = 0$, respectively:
\bea
   {\cal P}^{1/2}_{\hat \Phi} ({\bf k}, \eta)
   &=& {H \over 2 \pi} {\Gamma (\nu) \over \Gamma (3/2)}
       {1 \over a H |\eta|} \left( {k |\eta|\over 2} \right)^{3/2 -\nu}
       \Big| c_2 ({\bf k}) - c_1 ({\bf k}) \Big| {1 \over \sqrt{Q}},
   \label{P-spectrum-1} \\
   {\cal P}^{1/2}_{\hat \Phi} ({\bf k}, \eta)
   &=& {2 \sqrt{|\eta|} \over a} \left( {k \over 2 \pi} \right)^{3/2}
       \ln{(k|\eta|)} \times
       \Big| c_2 ({\bf k}) - c_1 ({\bf k}) \Big| {1 \over \sqrt{Q}}.
   \label{P-spectrum-2}
\eea
[For the gravitational wave, in order to get the correct numerical factor
we need to take into account of the two polarization states properly, 
see \cite{H-GW}.]
In (\ref{commutation}-\ref{P-spectrum-2}) the non-Einstein nature of the 
theory only appears in the parameter $Q$.
Although the simplest vacuum state with $c_2 = 1$ and $c_1 = 0$ is
often prefered in the literature, the power spectrums in
(\ref{P-spectrum-1},\ref{P-spectrum-2}) express the possible dependence
on the general vacuum state.

The condition used to get (\ref{Phi-k-sol}) may look special. 
However, as a matter of fact, it includes most of the proto-type inflation 
models investigated in the literature.
The exponential ($a \propto e^{Ht}$) and the power-law ($a \propto t^p$)
expansions realized in Einstein gravity with a minimally coupled 
scalar field lead to $\nu = {3 \over 2}$ and $\nu = {1 - 3p \over 2(1-p)}$, 
respectively \cite{QFT}.
The pole-like inflations ($a \propto |t_0 - t|^{-s}$) realized in 
the generalized gravities in (b)-(d) with the vanishing potential 
lead to $\nu = 0$;
these include the pre-big bang scenario based on the low energy effective 
action of the string theory \cite{pre-big-bang}.

As a perturbation scale reaches superhorizon size and the
perturbation evolution enters the classical regime
we can match the power spectrum in (\ref{P-spectrum-1},\ref{P-spectrum-2})
with the classical one based on the spatial averaging.
The later large scale perturbation evolution  is characterized by 
the conserved behavior of $\Phi$. 
Information about the classical structures can be recovered from $\Phi$ 
at the second horizon crossing epoch during the ordinary matter dominated 
era.  At this point it provides the initial conditions for the 
nonlinear evolution stage. 
This completes the connection between quantum fluctuations in the
early universe and large scale structures during the current epoch.
The exponential expansion and the large $p$ limit of the power-law expansion
lead to scale invariant spectra for density perturbations and 
gravitational waves which conform with observations, whereas the 
pole-like inflation models
based on the various generalized gravity with a vanishing potential 
lead to high power on small scales, see \cite{H-GW,Kin}.

\subsection*{6. Discussions}

In this paper we have presented a unified way of describing
the quantum generation and the classical evolution of scalar
structures and gravitational waves in a class of generalized gravity
theories. 
A rigorous treatment is made possible by two main ingredients:
first, the general conservation behavior on large scales 
(\ref{LS-sol}) which makes the classical evolution simple, 
and second, the exact solution in some generic expansion stages in 
(\ref{Phi-k-sol}) which makes the quantum perturbation generation simple.
One underlying reason for these simple results in generalized gravity
is the conformal equivalence between (\ref{Action-GGT}) and Einstein gravity 
with a minimally coupled scalar field, ignoring $L_m$ \cite{CT,H-CT}.
{}For nonvanishing $L_m$ two theories mathematically related by 
the conformal transformation would be physically different 
\cite{CT-interpretation}.

The gravitational action in (\ref{Action-GGT}) does not 
include a Ricci-curvature square 
term which naturally appears from one-loop quantum corrections, \cite{QFCS}.
Gravity with a Ricci-curvature square term does not have the conformal
transformation symmetry with Einstein gravity and may lead to different 
results for structure generation.
In particular, the recently popular string theory offers the possibility
that the observationally relevant structures leave the horizon near the
end of a pole-like inflation (pre-big bang) stage where the higher 
order string quantum correction terms are important.
Investigating structure generation processes in the strong quantum regime
is an interesting {\it open problem} which is left for the future endeavor.

\subsection*{Acknowledgments}

This work was supported in part by the KOSEF Grant No. 95-0702-04-01-3 and 
through the SRC program of SNU-CTP.


\end{document}